\documentclass[journal=jacsat,manuscript=article]{achemso}

\usepackage[version=3]{mhchem} 
\usepackage{svg}
\usepackage{tabularx}
\usepackage{subcaption, comment}
\usepackage{pdfpages}

\author{Ahmed Elrashidy}
\affiliation{Department of Physics, Astronomy, and Geosciences, Towson University, 8000 York, Towson, MD 21252, USA}
\email{aalras2@students.towson.edu}

\author{James Della-Giustina}
\affiliation{Department of Mathematics, Towson University, 8000 York, Towson, MD 21252, USA}

\author{Jia-An Yan}
\affiliation{Department of Physics, Astronomy, and Geosciences, Towson University, 8000 York, Towson, MD 21252, USA}
\email{jyan@towson.edu}

\title{Accelerated Data-Driven Discovery and Screening of Two-Dimensional Magnets Using Graph Neural Networks}

\abbreviations{IR,NMR,UV}

\begin{document}




\begin{abstract}
  In this study, we employ Graph Neural Networks (GNNs) to accelerate the discovery of novel 2D magnetic materials which have transformative potential in spintronics applications. Using data from the Materials Project database and the Computational 2D materials database (C2DB), we train three GNN architectures on a dataset of 1190 magnetic monolayers with energy above the convex hull ($E_{\text{hull}}$) less than 0.3 eV/atom. Our Crystal Diffusion Variational Auto Encoder (CDVAE) generates 11,100 candidate crystals. Subsequent training on two Atomistic Line Graph Neural Networks (ALIGNN) achieves a 93$\%$ accuracy in predicting magnetic monolayers and a mean average error of 0.039 eV/atom for $E_{\text{hull}}$ predictions. After narrowing down candidates based on magnetic likelihood and predicted energy, constraining the atom count in the monolayers to five or fewer, and performing dimensionality checks, we identify 190 candidates. These are validated using Density-Functional Theory (DFT) to confirm their magnetic and energetic favorability resulting in 167 magnetic monolayers with $E_{\text{hull}} < 0.3$ eV/atom and a total magnetization of $\geq$ $0.5 \mu_{B}$. Our methodology offers a way to accelerate exploring and predicting potential 2D magnetic materials, contributing to the ongoing computational and experimental efforts aimed at the discovery of new 2D magnets.
\end{abstract}

\section{Introduction}

Data-driven approaches have recently had a significant impact on discovering novel materials. 
Many methodologies in data-driven studies of materials utilize high-throughput DFT calculations, leading to the development of multiple databases of materials in one, two, and three dimensions. 
Prominent examples are the Materials Project (MP) Database,\cite{jain2013commentary} the Joint Automated Repository for Various Integrated Simulations (JARVIS),\cite{choudhary2020joint} the Open Quantum Materials Database (OQMD),\cite{OQMD1, OQMD2} the Computational 2D Materials Database (C2DB) \cite{haastrup2018computational,gjerding2021recent}, and the Computational 1D Materials Database (C1DB),\cite{moustafa2022computational} among others. 
More recently, the fusion of DFT and Machine Learning (ML) techniques has become an invaluable tool for discovering new materials and understanding the properties of existing ones. 
Specifically, the introduction of ML approaches has offset some need for the expensive computational resources that DFT calculations often demand.\cite{ml_materialscience} 
One notable example of the efficiency of this hybrid approach is the experimental realization of new ultra-incompressible materials guided through a combination of DFT calculations and Bayesian Optimization.\cite{zuo2021accelerating} 
This approach has also been expanded to utilize a stacked model that combines multiple ML models coupled with high-throughput DFT to discover Fe‐based bimetallic chalcogenides.\cite{pant2023dft}

High-throughput methods and ML models are also making progress in discovering and studying 2D materials. 
A combination of ML models and DFT calculations was used to explore the structural and thermodynamic stability of 2D materials, in addition to screening functional 2D materials for energy conversion and storage. \cite{manti2023exploring,schleder2019exploring,sorkun2020artificial} 
Moreover, a symmetry-based approach has been devised to screen for all thermodynamically stable combinations of binary and ternary 2D materials, and high-throughput DFT calculations were utilized in discovering 2D superconductors. \cite{wang2023symmetry,wines2023high}

Historically, the realization of a 2D magnet has been deemed infeasible due to the Mermin-Wagner theorem unless a magnetic anisotropy is present in the system. \cite{mermin1966absence} The first indication of the existence of intrinsic 2D magnetism down to the monolayer limit was demonstrated through the discovery of antiferromagnetic (AFM) order in FePS$_3$. \cite{lee2016ising} Following discoveries of ferromagnetic (FM) order in monolayer CrI$_3$, \cite{huang2017layer} Cr$_2$Ge$_2$Te$_6$,\cite{gong2017discovery} Fe$_3$GeTe$_2$,\cite{fei2018two} CrBr$_3$,\cite{zhang2019direct} and other materials have attracted much interest in this class of 2D materials. 

The discovery of new 2D magnets has become a very active field of research due to its potential in developing the next generation of nano-electronic devices encompassing spintronics applications.\cite{nano_2021} In fact, several experimental observations of room-temperature magnetism could mean that practical applications that capitalize on this class of materials may not be too far from becoming a reality. \cite{zhao2023room,wu2021strong,zhao2022two,zhang2022above,cai2022room,zhang2021room} Nonetheless, 2D magnets are mostly not air-stable which presents a challenge for adoption in potential applications. \cite{liu2023recent} Another important long-standing challenge is the realization of semiconducting magnets with critical temperatures above room temperature. \cite{lee2023possible} Additionally, many 2D magnets exhibit interesting quantum phenomena such as superconductivity,\cite{ontoso2023unconventional} spin liquid states,\cite{shaginyan2020theoretical,banerjee2016proximate} and topological phases.\cite{sachs2013ferromagnetic,han2019topological,khela2023laser,zhang2023magnetic} Consequently, this class of materials presents itself as a playground for quantum phenomena and strongly correlated effects in addition to its numerous potential applications. As a result, many studies have been devoted to high-throughput and ML investigations to discover new 2D magnets.

One of the notable successes of these studies is pointing out that Cr$_3$Te$_4$ is a promising candidate for two-dimensional magnetism.\cite{zhu2018systematic}. This material has been recently experimentally synthesized and verified to be a layer-dependent ferromagnet.\cite{wang2022layer} Additionally, high-throughput calculations have been utilized to screen for 2D magnetic materials based on experimental databases,\cite{torelli2020high} predict the ground state of collinear magnetic order,\cite{horton2019high} and search for magnetic topological 2D materials.\cite{choudhary2020computational} Other studies have also focused on the role of magnetic anisotropy in 2D magnets. \cite{minch2023data,torelli2019high,xie2021data}

Also, a semi-supervised ML approach was adopted to identify thermodynamically stable 2D magnets for spintronics applications.\cite{rhone2023artificial} 
Moreover, the investigation of 2D magnets by means of atomic substitution has been carried out for compounds of the forms A$_2$B$_2$Te$_6$  and A$^{i}$A$^{ii}$B$_4$X$_8$ using DFT and ML.\cite{rhone2020data,bhattarai2023investigating} 
ML approaches such as XGBoost have also proven to be effective in gaining a deeper physical understanding of the magnetic ordering of 2D magnets.\cite{chen2016xgboost,acosta2022machine} 

Despite the success of many high-throughput and ML techniques in predicting material properties and even discovering new ones, they are often limited in how they represent materials. This is due to the fact that crystalline materials often do not naturally lend themselves to traditional ML structures like vectors and matrices. To overcome this challenge, Graph Neural Networks (GNNs) emerge as a promising solution. 
Differing from traditional feed-forward multi-layer perceptron neural networks, GNNs excel in processing graph-structured input. \cite{GNN} 
Through message-passing mechanisms, GNNs can iteratively update node representations based on information from neighboring nodes and edges. This capability makes them particularly apt for material property exploration and inverse design.\cite{GNN_materials} 

Herein, we have generated new 2D magnets by using Crystal Diffusion Variational Autoencoder (CDVAE), a generative model designed for generating stable materials by learning from the data distribution of known materials.\cite{xie2021crystal} 
Additionally, we also employ the Atomistic Line Graph Neural Network (ALIGNN) \cite{choudhary2021atomistic} and the Materials Graph with 3-body Interactions neural Network (M3GNet).\cite{Chen2022}

The advantage of using diffusion models,\cite{sohl2015deep,song2019generative} in comparison to elemental substitution, is that it streamlines the material generation process. These models eliminate the need to wade through countless potential stoichiometries and symmetries, adopting a data-driven approach instead. 
Additionally, graph-based diffusion models have shown great promise in tasks like designing inorganic crystals,\cite{han2023design} generating 1D and 2D materials,\cite{lyngby2022data, Moustafa2023} producing potential superconducting materials,\cite{wines2023inverse} and curating stable spintronic materials.\cite{Siriwardane2023}

In this work, GNNs were utilized in conjunction with high-throughput DFT calculations to methodically develop an accelerated data-driven approach for 2D magnets discovery; this is achieved through a four-fold process. 
First, magnetic monolayers are generated by a CDVAE model. 
Then, the monolayers are screened through ALIGNN models in addition to charge neutrality, electronegativity, number of atoms, and dimensionality checks. 
Next, a symmetry-constrained relaxation is then performed using M3GNET IAP. 
Finally, further DFT relaxations and self-consistent calculations are performed to assess the magnetic properties and the thermodynamic stability of the screened monolayers.

\section{Computational Methods}
\subsection{CDVAE}

CDVAE generates new materials by learning from the available crystalline structures in the training data. 
The model does not directly use pre-defined chemical formulas; instead, it learns to generate crystals that are likely to be stable by identifying energy minimums and bonding preferences between different types of atoms from the data. Hence, the compositions, ratios of elements in the generated materials, and their chemical and physical properties are indirectly derived from the training data. 
Additionally, it incorporates stability through a noise conditional score network as the decoder of the Variational Autoencoder (VAE). \cite{kingma2013auto}
This decoder utilizes a harmonic force field to estimate the forces on atoms when their coordinates deviate from the equilibrium positions, thereby providing an important physical inductive bias for generating stable materials. 
CDVAE encodes permutation, translation, rotation, and periodic invariances through SE(3) equivariant GNNs adapted with periodicity. \cite{Batzner2022}
This allows CDVAE to generate valid, diverse, and realistic materials, in addition to its ability to optimize physical and chemical properties in latent space.

\begin{figure*}[tbp]
\centering
\includegraphics[scale=0.5]{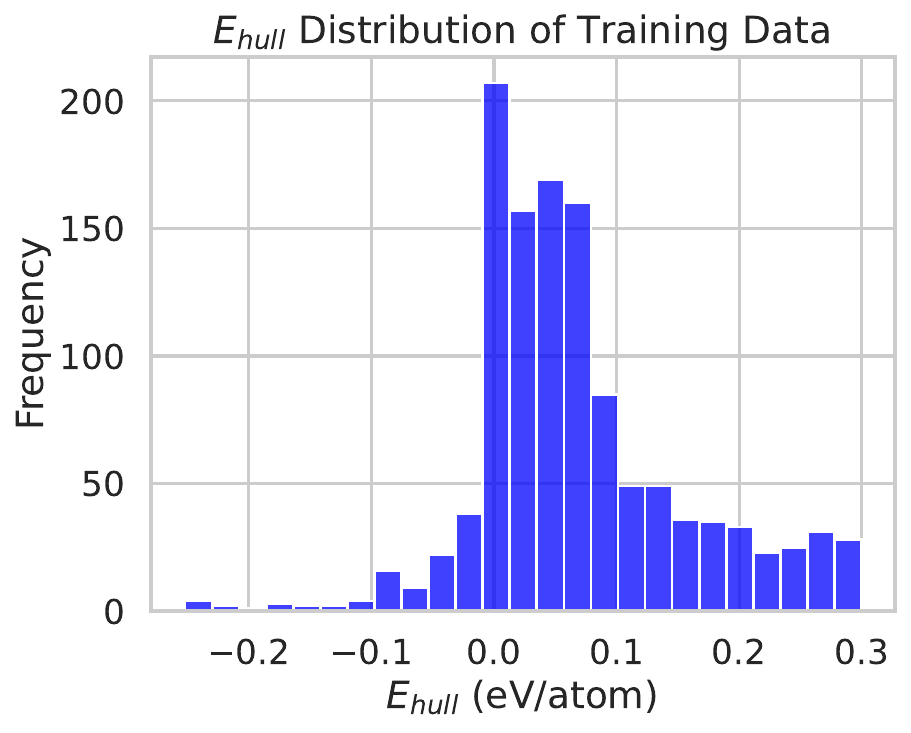}
\caption{Distribution of $E_{\text{hull}}$ of the 1190 training examples. The negative energies relative to the convex hull are due to monolayers that also exist in competing bulk phases with formation energies calculated using DFT+U which can lead to an overestimation of the formation energies. \cite{haastrup2018computational}} 
\label{ehull_training}
\end{figure*}

We utilized a CDVAE model to generate magnetic monolayers, training it on the available data of magnetic monolayers from C2DB. 
The database consists of more than 15,000 monolayers with a multitude of properties calculated through DFT.
From C2DB, we curated a dataset of 1190 monolayers by filtering only magnetic monolayers with $E_{\text{hull}}$ less than 0.3 eV/atom. 
The dataset was partitioned using an 80-10-10 split, allocating 80\% for training, 10\% for validation, and 10\% for testing. 
It should be emphasized that the $E_{\text{hull}}$ cutoff value applied in this context serves not as a stability criterion, but simply as a threshold for selecting training data.
This value of 0.3 eV/atom was chosen to be consistent with previous work that employed CDVAE to incorporate new entries to C2DB. \cite{lyngby2022data} 
The distribution of the energy relative to the convex hull of the training set is shown in Figure~\ref{ehull_training}. 

\subsection{ALIGNN}
This model leverages both the atomic graph structure of the material and its corresponding line graph, which is derived from the original graph by transforming each edge into a vertex. The set of edges in the line graph is then constructed based on whether two edges shared a mutual vertex in the original graph; if so, they now share an edge in the line graph. 
By associating edges with the interatomic distances and vertices with the elemental properties of each atom in the original graph, alongside correlating edges with bond angles and vertices with interatomic bonds in the line graph, ALIGNN is able to capture an extensively higher degree of information. 
This is achieved by sequentially performing message passing on each graph and facilitating information propagation between the two. 

We trained two ALIGNN architectures using a dataset of 154,718 materials from the materials project (MP) database and an 80-10-10 split for training. 
The first model is a regression model to predict the $E_{\text{hull}}$ of the entries, and the second is a classification model to predict magnetism.

\subsection{M3GNET Relaxation}

M3GNet is a graph neural network architecture designed to be a universal (interatomic potential) IAP for materials. 
It is trained on a large dataset of energies, forces, and stresses from structural relaxations performed by the Materials Project, and is capable of predicting the properties of individual atoms as well as the overall crystal structure. 
M3GNet uses a three-body interaction model and is able to capture long-range interactions without the need to increase the cut-off radius for bond construction. 
The architecture represents the elemental information for each atom as a learnable embedding vector, making it readily extendable to multicomponent chemistries.

Force Field methods have been suggested as pre-structure-optimizers before performing DFT calculations. \cite{wines2023recent} 
By utilizing the default pre-trained M3GNet IAP, we conducted a symmetry-constrained relaxation of the cell shape, which is achieved by applying tensile and compressive strains within the plane of the monolayers. 
This allows us to approach the neighborhood of the potential energy surface (PES) minima at a fraction of the cost of DFT before performing full relaxations.

\subsection{DFT Calculations}

For the DFT calculations, we employed the projected augmented wave (PAW) method as implemented in the Vienna \textit{ab initio} Simulation Package (VASP).\cite{kresse1996efficient,kresse1999ultrasoft} 
We chose the Perdew-Burke-Ernzerhof (PBE) \cite{perdew1996generalized} formulation for the generalized-gradient exchange-correlation functional (GGA). 
The Brillouin zone underwent sampling using a $9\times9\times1~~k$-point grid mesh \cite{monkhorst1976special} with a 600 eV plane wave cutoff energy. 
This mesh size was chosen so that after M3GNET relaxations, the k-point density for the smallest lattice vector in the considered materials is 4 \text{points}/Å$^{-1}$.

Limiting our considerations to the ferromagnetic states, initial magnetic moments were assigned to the atoms for collinear magnetic calculations without spin-orbit coupling (SOC). 
Transition metal elements were initialized with a magnetic moment of $6 \mu_{B}$, while all other elements were initialized to a value of $0.5 \mu_{B}$. 
This intentional overestimation aims to facilitate convergence and relaxation into the accurate ferromagnetic ground state.

Each structure was fully relaxed until the Hellmann-Feynman forces on every atom were less than $2 \times 10^{-2}\ \text{eV/Å}$, and the energy convergence criterion reached $10^{-6}\ \text{eV}$. 
For self-consistent evaluations, a more stringent energy convergence criterion of $10^{-7}\ \text{eV}$ was applied.

It is worth noting that this approach does not consider AFM order and therefore could miss energetically favorable AFM phases, if present. 
As this work is concerned with developing an approach for accelerating 2D magnet discovery, considering AFM states that typically require increasing the size of the unit cell goes beyond our scope.
Additionally, we have used PBE functionals throughout with no inclusion of a Hubbard-U correction in order to be consistent with the computational methodology of C2DB; this also allows for more consistent energetic favorability analysis. 
Nonetheless, many magnetic 2D materials exhibit strongly correlated effects and are better suited for more advanced methods that account for correlation in the outer electronic shells such as DFT+U,\cite{anisimov1991band, ylvisaker2009anisotropy} dynamical mean field theory (DMFT),\cite{georges1996dynamical} and Quantum Monte Carlo.\cite{foulkes2001quantum} 
These methods have proven to be more accurate in describing materials manifesting strong correlation effects. \cite{karp2021dependence,zang2022dynamical,jiang2022monte,wines2023quantum,wines2023systematic} 

\section{Results \& Discussion}
\subsection{Generating 2D Magnets}

In C2DB, the material properties of the greatest importance for our goal are the magnetic and thermodynamic ones.
Specifically, we are interested in the energy above the convex hull ($E_{\text{hull}}$). 
The energy of the convex hull is derived from the Gibbs free energy at 0 K temperature and 0 atm pressure by constructing the convex hull set of the normalized formation energy with respect to the number of atoms.
The energy above hull calculation provides an additional advantage over using the formation energy per atom ($E_f$).
This advantage stems from $E_{\text{hull}}$ assessing the energetic favorability of compounds relative to other competing phases in terms of their formation energies.
Therefore, a compound's stability is inversely proportional to its energy above the convex hull; the less stable it is, the higher its $E_{\text{hull}}$.
In many data-driven approaches to generating materials, $E_{\text{hull}}$ has become a standard measure to assess the thermodynamic stability of the generated materials.\cite{wines2023inverse,wang2023symmetry,lyngby2022data,priya2021accelerated,choubisa2023interpretable,zhao2023physics,Siriwardane2023,Moustafa2023}

\begin{figure*}[h]
\centering
\includegraphics[scale=0.5]{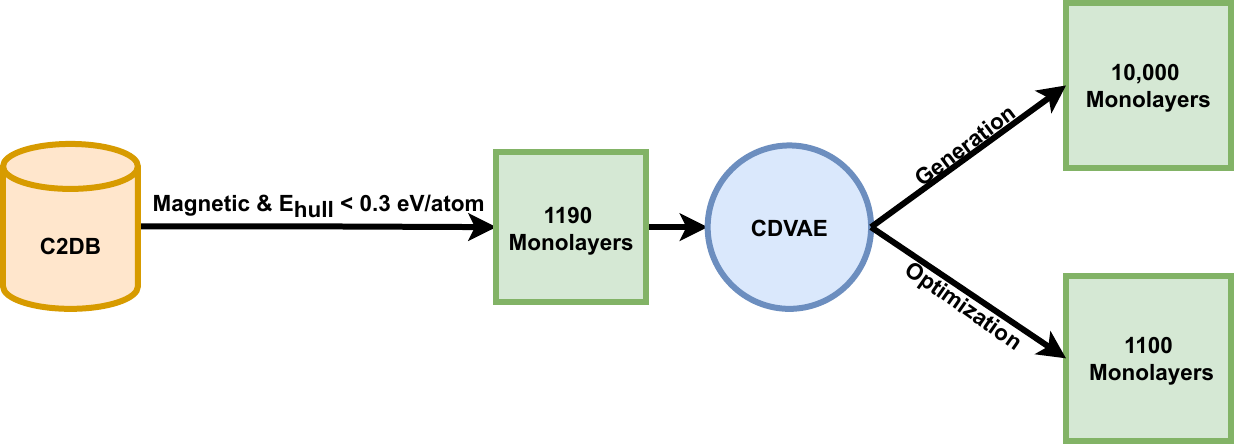}
\caption{Outline of training CDVAE to generate 11,100 candidate monolayers for further screening.} 
\label{CDVAE}
\end{figure*}

\begin{figure*}[h]
\centering
\includegraphics[scale=0.4]{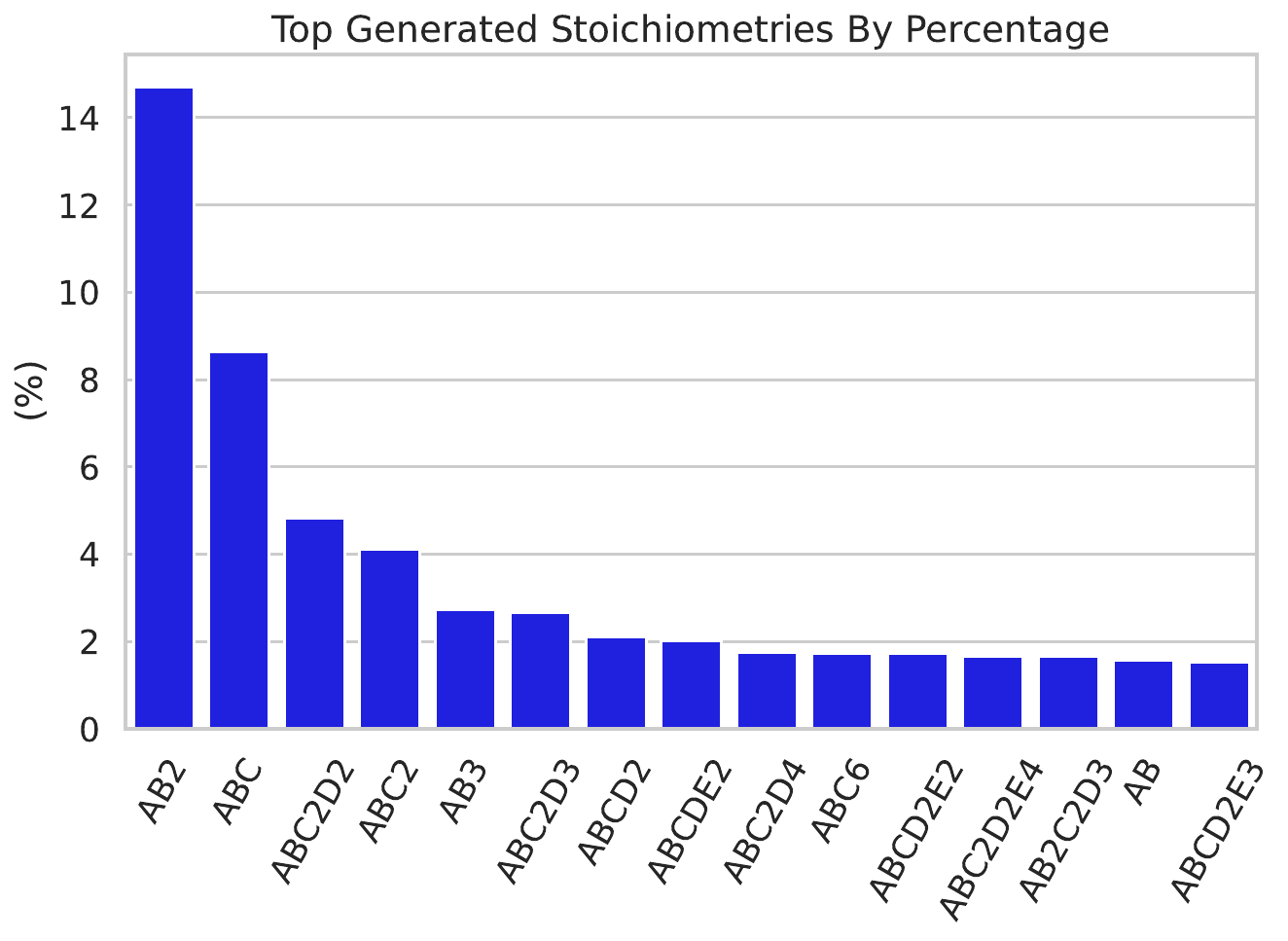}
\caption{The distribution of stoichiometries generated by CDVAE.} 
\label{CDVAE_stoich}
\end{figure*}

We trained the CDVAE model to perform two tasks, generating monolayers both with and without optimization for the energy above the convex hull in latent space. 
The model yielded 11,100 candidate monolayers, which were subsequently screened. This process to generate these monolayers is depicted in Figure~\ref{CDVAE}. For reference, the distribution of the top-generated stoichiometries by CDVAE is shown in Figure~\ref{CDVAE_stoich}.

The larger portion of the materials were generated without optimization of the $E_{\text{hull}}$ in the latent space. 
This was done to increase the number of candidates since the training dataset was filtered to only contain materials within proximity to $E_{\text{hull}}$.

\subsection{Training ALIGNN Discriminator Models}

While DFT remains the gold standard in determining the properties of 2D materials, including thermal stability and magnetic properties, performing thousands of DFT calculations requires a considerable amount of time and computational power. 

\begin{figure*}[h]
\centering
\includegraphics[scale=0.5]{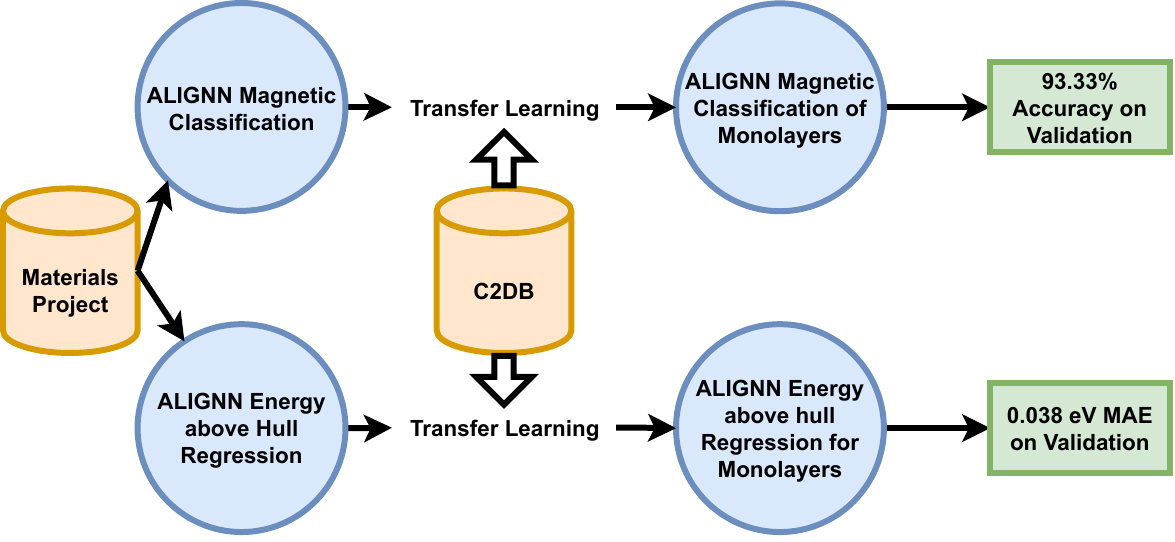}
\caption{Outline of training CDVAE to generate 11,100 candidate monolayers for further screening.} 
\label{ALIGNN}
\end{figure*}

This becomes a crucial factor when considering that many of the generated monolayers may not be magnetic or thermodynamically stable. 
Hence, we utilize the ALIGNN models to filter candidates by their magnetic probability and thermal stability before running DFT calculations.
The overall outline of how the ALIGNN models are utilized and the results of training are illustrated in Figure~\ref{ALIGNN}.

The regression model predicts $E_{\text{hull}}$ and achieves an MAE (Mean Average Error) of 0.028 eV/atom. 
We then employ transfer learning and further train the model on the monolayers in C2DB to predict the energy above the convex hull and obtain an MAE of 0.039 eV/atom. 
The second model is utilized to classify whether a material is magnetic, which 
has previously been accomplished in ALIGNN performance benchmarks by specifying a cutoff value of 0.05 $\mu_{B}$ for the total magnetic moment. \cite{choudhary2021atomistic} 
Using this strategy, the model attains an accuracy of 87.30\% on the test dataset, likely due to the fact that this approach effectively classifies the data into ferromagnets or non-ferromagnets. 
Moreover, within the same material, structural deformations through strain may be associated with a change in the magnetic order. \cite{carpenter2012magnetoelastic, webster2018strain, Hu2020} 
Fortunately, the Materials Project API provides a classification of whether a material is magnetic by considering multiple states of magnetism including ferromagnetism, antiferromagnetism, and other magnetic phases. 
By training the model using this classification label instead of using a cutoff value for the magnetic moment, the model accuracy improves and reaches 93.33\%. 
Again, we apply transfer learning to train the model on classifying the monolayers in C2DB into magnetic or non-magnetic, reaching an accuracy of 93.33\% on C2DB data. 
The performance of both the classification and regression models with and without the use of transfer learning is compared in Table~\ref{table1}. 
An early stopping condition was employed, set to activate after 15 epochs without performance improvement on the validation data. 
As evidenced by Table~\ref{table1}, models utilizing transfer learning met this condition more quickly, but interestingly did not improve the accuracy score. 
Additionally, the $E_{\text{hull}}$ regression model achieves a better performance in fewer epochs.

\begin{table}[tbp]
\centering
\caption{The effect of transfer learning on performance on the test dataset and the speed of training.}
\label{table1}
\begin{tabular}{|c|c|c|c|c|}
\hline
& \multicolumn{2}{c|}{Magnetic Classification} & \multicolumn{2}{c|}{$E_{\text{hull}}$ Regression} \\
\cline{2-5}
& Training Epochs & Accuracy & Training Epochs & MAE \\
\hline
With Transfer Learning & 18 & 93.33 & 159 & 0.044 \\
\hline
Without Transfer Learning & 32 & 93.33 & 169 & 0.039 \\
\hline
\end{tabular}
\end{table}

\begin{figure}[htbp]
\includegraphics[scale = 0.55]{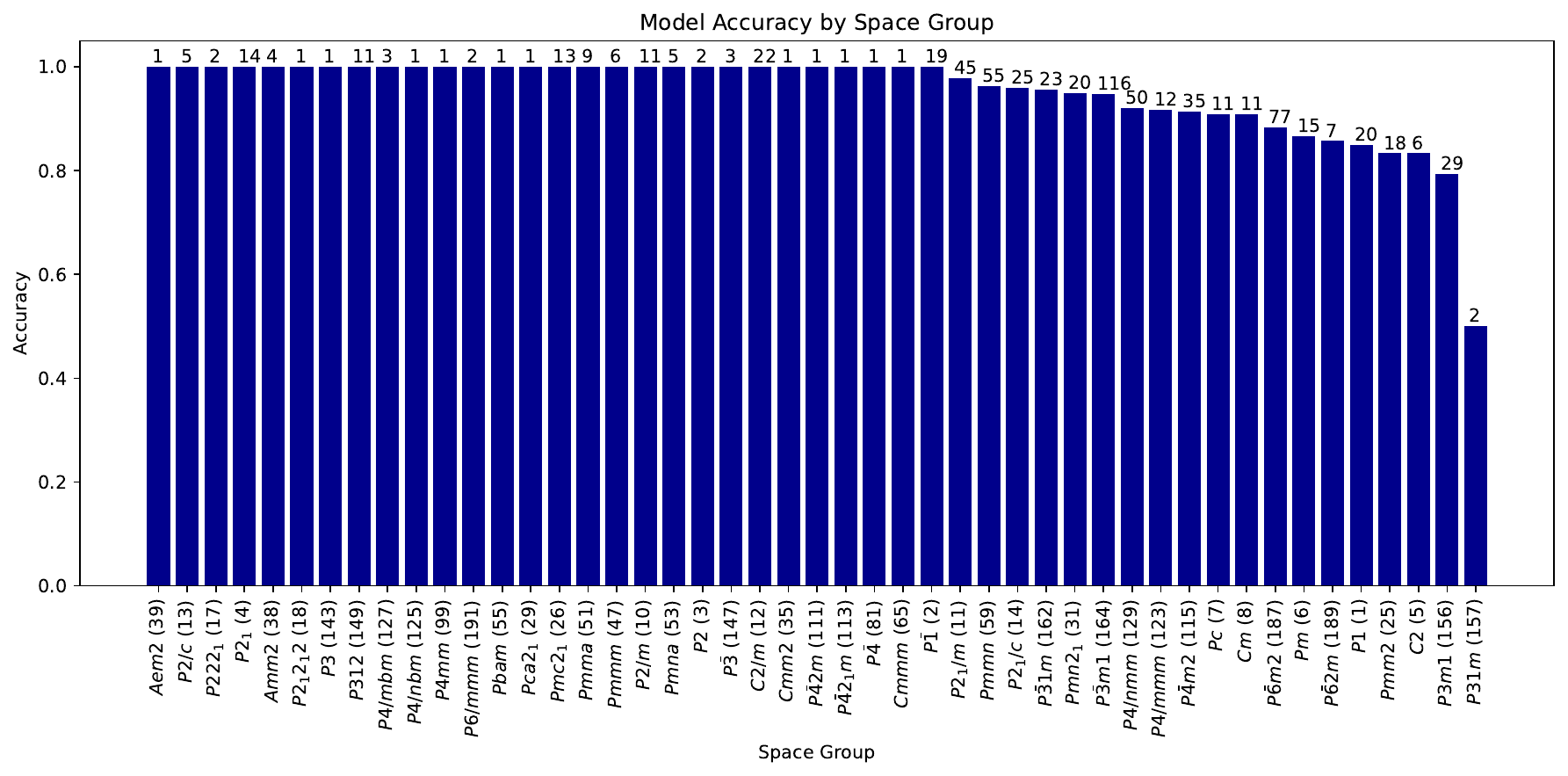}
\caption{Model accuracy across different space groups in the test data; the number of test data points from each group is indicated on each bin.}
\label{groups}
\end{figure}

To demonstrate the model's robust performance in classifying magnetic materials across different symmetries, we present the model's accuracy among the space groups in the testing data in Figure~\ref{groups}. 
As depicted by the Figure, the model accuracy is high across the majority of the space groups in the test data. 

\subsection{Lattice Relaxation Through M3GNET IAP}
\begin{figure}[H]
\includegraphics[scale = 0.4]{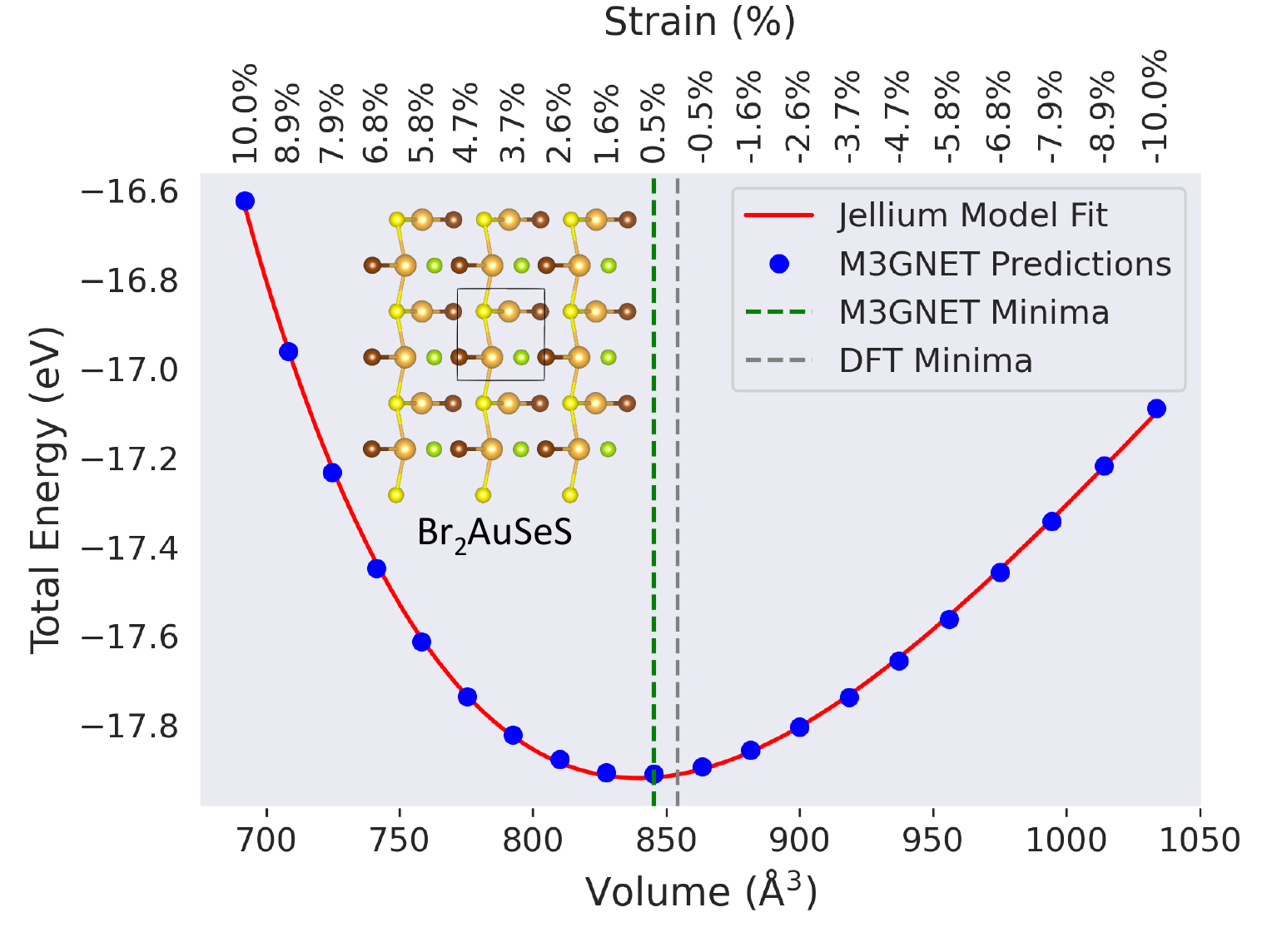}
\caption{Symmetry-constrained relaxation by applying strain fitted to the jellium model. The energetic minima obtained using DFT and M3GNET are indicated in the plot. }
\label{relax}
\end{figure}

The M3GNet IAP is a graph neural network-based approach using interatomic potentials that predicts the energies, forces, and stresses of a material \cite{Chen2022}. 
To find the energetic minima, we use this model to relax the crystal shape by applying in-plane strain on the lattice. 
Then, the predicted energies are fitted to the volumes using the stabilized jellium equation of state provided through the atomic simulation environment (ASE).\cite{larsen2017atomic,alchagirov2003reply} 
A visualization of the fitting process, as strain is exerted on a monolayer of Au$_3$Br$_2$SSe from C2DB (ID = 4605), is presented in Figure~\ref{relax}.

To evaluate the efficacy of this model at performing symmetry-constrained relaxation, we employed the 2-norm difference between the symmetry-constrained IAP relaxation and DFT relaxed lattice vector matrices. 
For a given material, this is computed as \( ||\mathbf{A}_{\text{IAP}} - \mathbf{A}_{\text{DFT}}||_2 \), where \( \mathbf{A}_{\text{IAP}} \) and \( \mathbf{A}_{\text{DFT}} \) are the respective lattice vector matrices. 
For a given set of materials, the average 2-norm difference is given by \( \bar{D} = \frac{1}{N} \sum_{i=1}^N ||\mathbf{A}_{\text{IAP},i} - \mathbf{A}_{\text{DFT},i}||_2 \), where \( N \) is the total number of materials. 
This average \( \bar{D} \) provides a statistical measure of the IAP method's accuracy in reproducing DFT-computed lattice parameters across a variety of materials. 
By computing the average 2-norm difference across all entries in C2DB, we obtained a value of  \( \bar{D} \) = 0.19 Å. 
This result demonstrates that, through the utilization of machine learning, M3GNET IAPs are capable of achieving near DFT performance levels at a significantly reduced computational cost. 
For comparison, the default pre-trained atomistic line graph neural network-based force fields (ALIGNN-FF) model achieves \( \bar{D} \) = 0.47 Å. 
The result of the symmetry-constrained relaxation fit using ALIGNN-FF is shown in the supplementary in Figure S2.

\subsection{Screening and DFT calculations}

\begin{figure}[H]
\includegraphics[scale = 0.4]{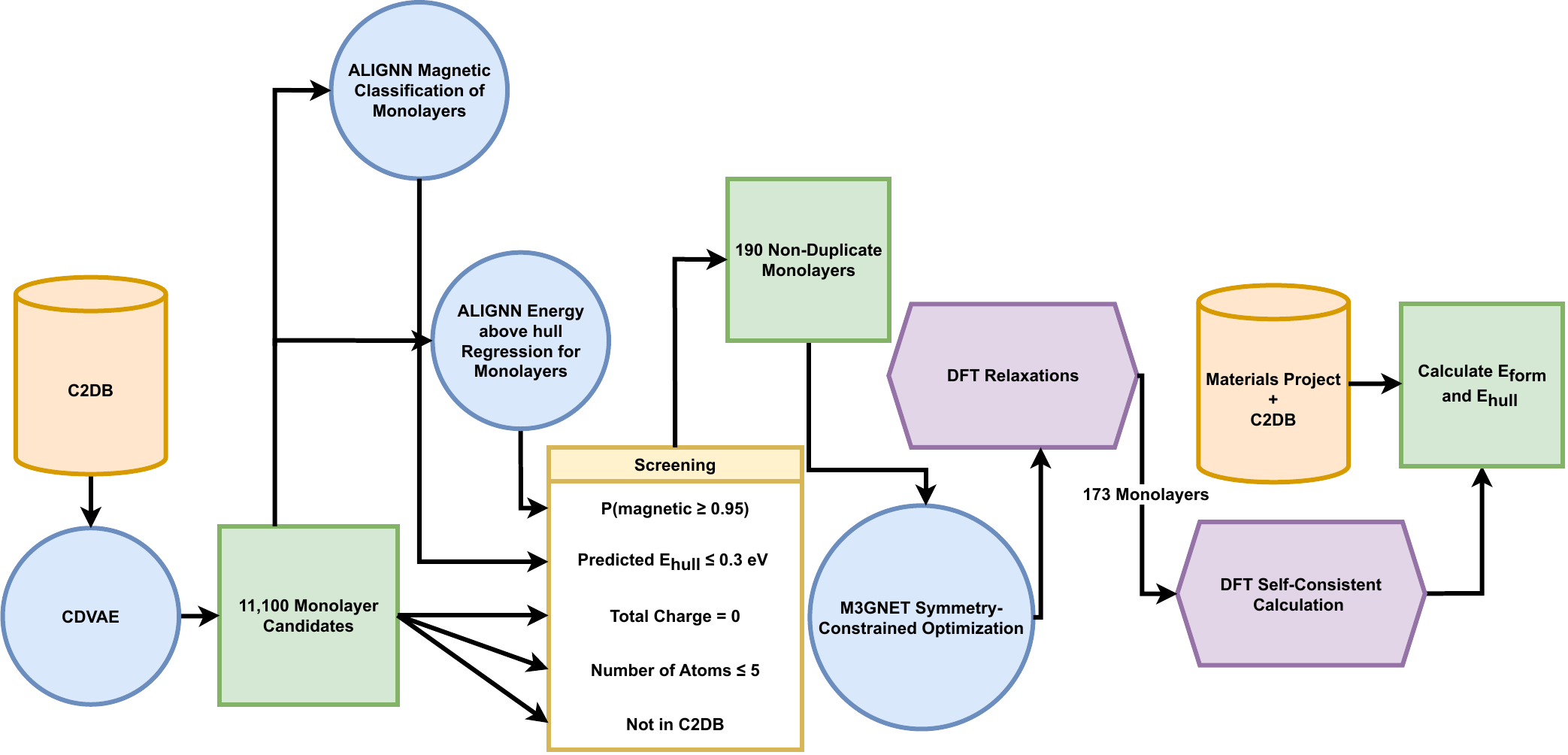}
\caption{The full data pipeline of material generation, screening, and DFT calculations.}
\label{pipe}
\end{figure}

Now that all the necessary components for generating, screening, and symmetry relaxation have been developed, we integrate them into a single pipeline, shown in Figure~\ref{pipe}.

After the monolayer candidates have been generated by CDVAE, they are passed through a screening step. 
The trained ALIGNN discriminator models are then used to make predictions on the generated monolayers. 
We only keep monolayers where the classification model outputs a 95\% probability of being magnetic and the regression model predicts the value of $E_{\text{hull}} \leq$  0.3 eV/atom. 
We also check for charge neutrality and run electronegativity tests as implemented in the Semiconducting Materials from Analogy and Chemical Theory (SMACT) Python library.\cite{davies2016computational}

Additionally, by constraining the number of atoms in the crystal to 5 or less, we further refined the candidate pool to an order of hundreds.
This value was chosen because it loosely aligns with the 25-percentile value of the number of atoms in the magnetic monolayers in C2DB, shown in Figure S1. 
Using a smaller number of atoms allows us to perform less computationally expensive calculations, given that DFT methods typically scale as $\mathcal{O} (N^3)$ where $N$ is the number of atoms. \cite{sholl2022density} 
Duplicate materials were removed by comparing the stoichiometries to the entries available in C2DB. 
Finally, we implement a dimensionality check step based on covalently bonded clusters to determine the dimensions of connected subunits in a structure, implemented in the Python materials genomics (Pymatgen) library. \cite{ong2013python,cheon2017data}

After screening, we end up with 190 non-duplicate candidates compared to C2DB. 
This relatively large number of candidates highlights CDVAE's ability to generate new materials, especially when considering that more than 3000 materials in the database have been added to the database by a CDVAE model. 
This potentially indicates CDVAE models are capable of generating new materials from unexplored chemical spaces by simply adjusting the training data to include examples that are representative of the desired properties. 

Additionally, only four of these materials out of the 1100 materials were generated through optimization for the energy above the convex hull in latent space. 
All other materials were generated without optimization for any properties in latent space. 
It may be that letting the diffusion model explore the chemical space without trying to optimize for a property in latent space could allow it to generate more diverse materials. 
Since materials in the training data have low energies above the convex hull, this property is inherited in the generated crystals. 
Interestingly, the thousands of materials added to the C2DB database by a CDVAE model were generated without any optimization of energies above the convex hull.

Proceeding with our workflow, we performed IAP symmetry-constrained optimizations of these crystals followed by DFT relaxations and self-consistent calculations. 
The relaxed structures are then further screened through four different, independent tests to exclude materials that did not relax into two-dimensional structures. \cite{pan2021benchmarking,gorai2016computational,larsen2019definition} 
In this step, a structure is considered to be two-dimensional only if all four dimensionality tests have been passed. 
This reduces down the total number of materials considered in this work to a final count of 173. 

After checking C2DB, Materials Project database, JARVIS, and OQMD for the presence of 
generated materials, 109 were deemed to be unreported as bulk or two-dimensional with 64 known materials found in OQMD. And all the duplicates we found in JARVIS and the materials project database were also included in OQMD.

 The formation energy per atom, \(E_{f}\), for each structure was computed using the formula: \(E_{f} = \frac{E_{\text{total}} - \sum_i n_i E_i}{N}\), where \(E_{\text{total}}\) is the total energy of the structure (obtained for DFT), \(n_i\) is the number of atoms of element \(i\) in the structure, \(E_i\) is the energy of element \(i\) obtained from the Materials Project database, and \(N\) is the total number of atoms in the structure. 
 
\begin{figure}[H]
\includegraphics[scale = 0.5]{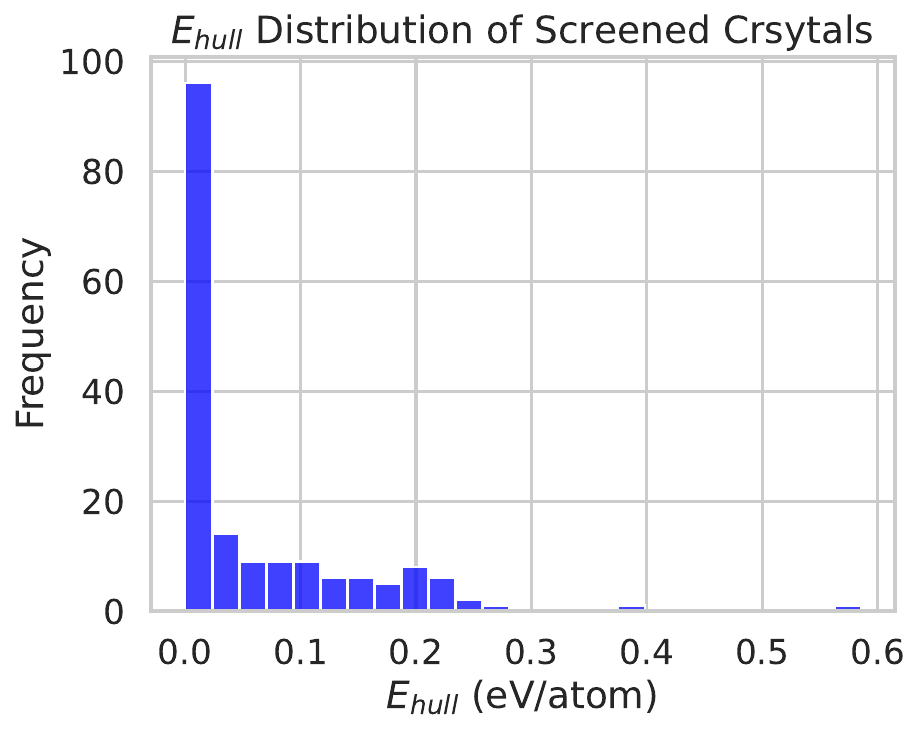}
\caption{The distribution of $E_{\text{hull}}$ of 173 final monolayers generated by the pipeline.}
\label{ehull-cdvae}
\end{figure}

In C2DB, the energy above the convex hull is constructed by considering the formation energy of elementary, binary, and ternary bulk crystals in OQMD. \cite{haastrup2018computational,gjerding2021recent} 
Naturally, this limits the number of considered phases and excludes 2D materials from competing phases. 
To expand the scope of the considered competing phases, we have computed the energy above the convex hull using the formation energies of compounds available in the Materials Project database and C2DB without any restriction on the number of constituent elements. 
For competing phases in the Materials Project database, only entries computed with GGA functionals and without the inclusion of a Hubbard U parameter were considered to ensure the exclusion of any energetic inconsistencies that may arise from employing different U values. 
The distribution of $E_{\text{hull}}$ of the monolayers generated through this pipeline are shown in Figure~\ref{ehull-cdvae}. 
Of the generated monolayers, 91 achieved an $E_{\text{hull}}$ of 0, while 171 displayed an $E_{\text{hull}} \leq 0.3$ eV/atom.  
It is worth noting, that only two materials were found to have $E_{\text{hull}} > 0.3$. 

While the energy above the convex hull provides a theoretical assessment of thermodynamic stability, materials with energies above the convex hull may be experimentally realizable for many reasons. 
First, PBE functionals deviate from experimental values of formation energies by $\approx$ 0.25 ev/atom, which in turn affects the accuracy of the energy above the convex hull predictions. \cite{pandey2015heats} 
Secondly, the monolayers generated herein are being proposed as freestanding monolayers and ignore external interactions with the environment such as substrate effects that may stabilize these monolayers. 
For example, silicene has a formation energy of 0.66 eV/atom \cite{haastrup2018computational} but can be synthesized on a metal substrate. \cite{malyi2019energy,li2018epitaxial} 
Moreover, even materials such as MoS$_2$ and ZrS$_2$ with $E_{\text{hull}} > 0$ have been experimentally synthesized. \cite{malyi2019energy,haastrup2018computational} 
In fact, based on the hull energies of 55 experimentally observed 2D materials, the criterion for high thermodynamical stability of 2D materials was proposed to be $E_{\text{hull}}$ $\leq$ 0.2 eV/atom \cite{haastrup2018computational} for calculations performed using PBE functionals. 
This criterion has been adopted in first-principle calculations to perform a screening of 2D materials suitable for photoelectrocatalytic water splitting. \cite{schleder2019exploring}

\begin{figure*}[tbp]
\centering
\includegraphics[width=1.00\textwidth]{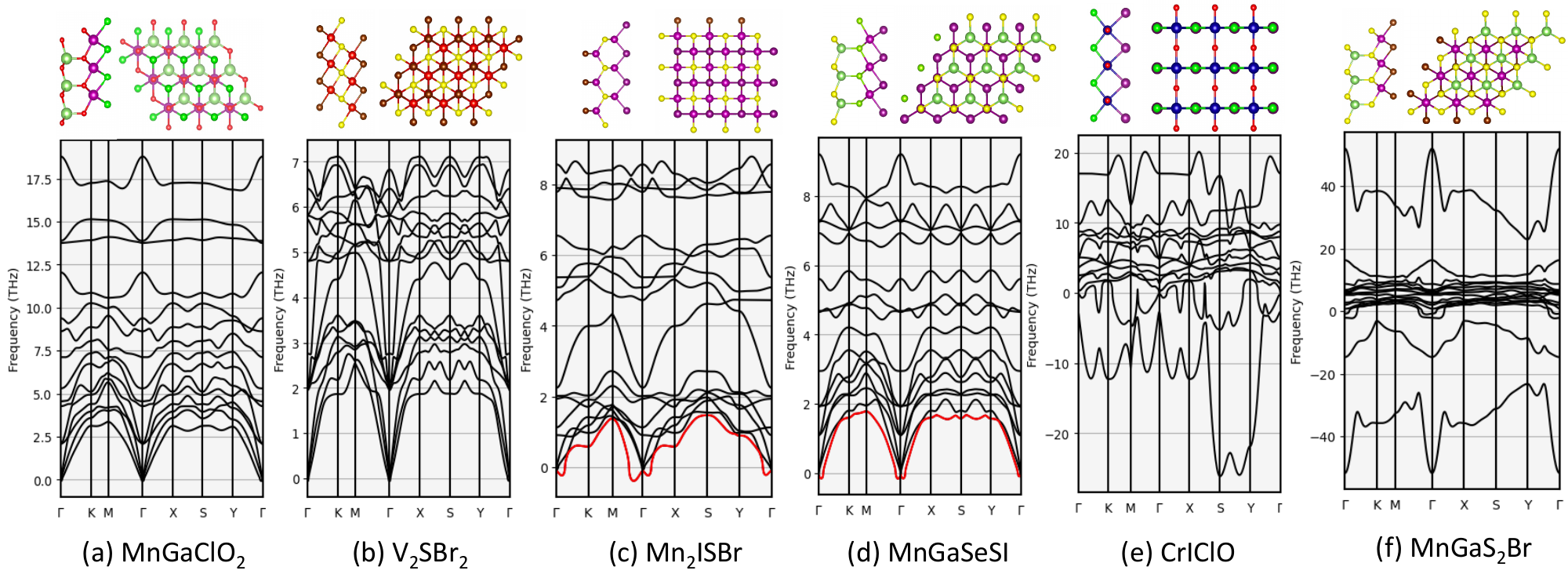}
\caption{Phonon band dispersions of 6 materials generated through our workflow; (a) and (b) are dynamically stable, (c) and (d) are likely stable, while (e) and (f) are likely unstable. A top and side view of the generated crystals are shown atop each subfigure to the right and left, respectively.} 
\label{phonon}
\end{figure*}

In addition to thermodynamic stability, mechanical and dynamical stability are also necessary for the realization of 2D materials. 
Dynamical stability can be assessed through the absence of imaginary modes in the phonon band dispersions of materials. 
We used the small-displacements method, implemented in the Phonopy software package \cite{togo2023implementation}, to calculate the phonon band dispersions of the 6 materials shown in Figure~\ref{phonon}. 
These materials have been confirmed to be magnetic and $E_{\text{hull}}$ = 0. 
Two of these materials, MnGaClO$_2$~\&~V$_2$SBr$_2$, display no imaginary phonon modes and therefore are dynamically stable. 
The phonon band dispersions of Mn$_2$ISBr~\&~MnGaSeSI show slightly imaginary phonon modes near the $\Gamma$ point, but are likely dynamically stable. 
Such slight imaginary modes have been reported in DFT-calculated phonon dispersions of many 2D materials, including experimentally realized ones such as GaSe and have been attributed to interactions between periodic monolayer images and numerical instabilities. \cite{radescu2019origin,ataca2012stable,zheng2015monolayer,zolyomi2014electrons} These small imaginary modes occur at the long-wavelength limit, typically in the acoustic branch possessing the lowest frequency near the $\Gamma$ point; we have highlighted this acoustic phonon branch in red for both materials. 
To the best of our knowledge, these four likely stable crystals have not been reported in bulk or as monolayers in any database at the time of writing.

The last two crystals of MnSeS and CrIClO exhibit large imaginary phonon modes and are likely unstable. 
Large negative phonon modes typically indicate dynamical instabilities in the structure. 
However, they may also indicate that the crystal is on the verge of structural transitions or distortions, such as the formation of phases hosting a Charge Density Wave (CDW).\cite{gruner2018density} 
For example, MoS$_2$ undergoes a transition from 1T to 1T' through a 2$\times$1 distorted structure. \cite{qian2014quantum, liu2018phase} 
Furthermore, many transition metal dichalcogenides (TMDs) monolayers undergo a transition to a CDW including NbSe$_2$,\cite{xi2015strongly} TaS$_2$,\cite{yang2018enhanced} TaSe$_2$,\cite{ryu2018persistent} TiSe$_2$,\cite{sugawara2016unconventional} VS$_2$,\cite{van2021full} and NbTe$_2$.\cite{bai2023realization} 
Moreover, a transition to a CDW has been theoretically predicted for the CrTe$_2$ monolayer and was shown to stabilize the imaginary phonon modes through crystal by a transition to a $\sqrt{3} \times \sqrt{3}$ supercell. \cite{otero2020controlled} 
In fact, because these transitions typically require constructing supercells capable of accommodating the onset of the new phase, one of the approaches to stabilizing such materials is to extend the atoms along the direction of instability. 
This approach has been utilized as the basis for a high-throughput DFT study to stabilize monolayers that exhibit structural instabilities in C2DB. \cite{manti2023exploring} 
Finally, accounting for strong electron correlations through a Hubbard U as low as 1 eV has been found to stabilize the imaginary phonon modes in 1T-CrTe$_2$. \cite{liu2022structural} 
However, considering the aforementioned approaches to stabilize the materials generated here goes beyond the scope of this work; our aim is to provide motivation and a blueprint for accelerating the discovery and screening of materials with specified properties.

\begin{figure}
    \centering
    \includegraphics[scale=.36]{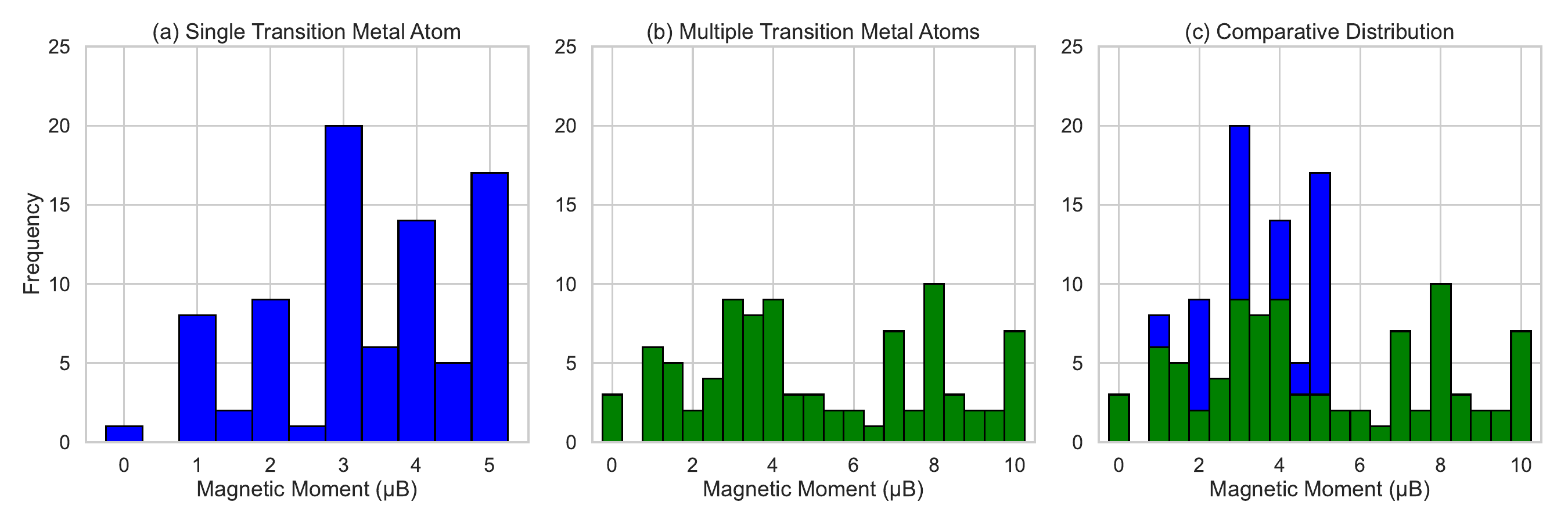}
    \caption{(a) The distribution of the total magnetic moment when a single transition metal atom exists shows peaks at integer values. (b) The distribution of magnetic moments with more than one transient metal atom exhibits a wide range of values. (c) A combined histogram displaying both magnetic moment distributions.}
    \label{fig:magneticmoments}
\end{figure}

Of the materials analyzed, only two demonstrated a total magnetic moment less than $0.5 \mu_{B}$. 
On average, materials exhibited a magnetic moment of $4.21 \mu_{B}$. 
The distribution of the DFT calculated magnetic moments is shown in Figure~\ref{fig:magneticmoments}. 
In subfigure (a), we have binned the total magnetic moments of crystals containing only a single transition metal atom. 
Clearly, the magnetic moments from the screened structures prominently exhibit peaks at integer values, suggesting a strong correlation with atomic electronic configurations. 
Notably, while peaks can be observed at the bins representing integer values, any data points are absent at $6\mu_{B}$ and $7\mu_{B}$ since configurations with 6 or 7 unpaired electrons are not achievable in a true ground state. 
As crystals with more magnetic moments are considered (b), we observe an increase of the total magnetization to $ \approx 5 \mu_{B}$ and climbing to $ \approx 10 \mu_{B}$. 
Because the considered materials contain at most two transition metal atoms, these large values have likely been produced by two magnetic atoms, each possessing a magnetic moment of $ \approx 5 \mu_{B}$.

The pattern largely aligns with the behavior of individual magnetic atoms. 
This consistency is observed in the majority of the screened structures, where we constrained the number of atoms to five or less, and most compounds featured just one magnetic element. 
This dataset accentuates the intricate relationship between atomic configurations and resultant magnetic moments, particularly when compounds comprise multiple magnetic atoms.

Overall, from the 190 materials we performed DFT calculations with, 173 were truly two-dimensional. Of those 173, two materials possessed $E_{\text{hull}}$ $>$ 0.3 eV/atom and 4 materials displayed a total magnetic moment of less than $0.5\mu_{B}$ 
This demonstrates the CDVAE's capability to generate realistic structures and underscores the efficiency of the discriminator ALIGNN models in learning the underlying distributions of the energy above the convex hull and magnetism. 

We have made the monolayer data available at \url{(https://github.com/rashigeek/2D-mag-GNN} including the structure of all 173 monolayers. The GitHub repository also includes the code used to produce the results. Additionally, we have provided a complete table of each material, its magnetic moment, and the energy above the convex hull in the supplementary material.

\subsection{Conclusions}

In this study, we presented an integrated computational approach that combines deep learning, symmetry-constrained optimizations, and DFT calculations to generate and screen potential 2D magnetic materials. 
The process begins with the generation of candidate monolayers using the CDVAE model. 
Candidates are then rigorously screened based on magnetic probability and thermal stability predictions from trained ALIGNN discriminator models. 
Screened candidates undergo lattice relaxation using the M3GNet IAP, with final properties computed using DFT.

Our findings demonstrate that the proposed pipeline is effective in predicting magnetic thermodynamically stable monolayers. 
The observed distributions of magnetic moments and energies above the convex hull provide a deeper understanding of the interplay between atomic configurations and magnetic properties, revealing distinct patterns and behaviors. 
By integrating ML and traditional computational techniques, we can expedite the discovery process, paving the way for next-generation materials with properties tailored to specific applications. 
Moreover, due to the multitude of properties that can also be analyzed using a similar approach to what we have discussed, our work presents a guideline on how to accelerate inverse-design material generation for optimized properties.

\begin{acknowledgement}

The authors thank Kristian Sommer Thygesen for kindly providing the C2DB database. Ahmed E. would also like to thank Ryan Paxson for valuable discussions. 

This work used EXPANSE at the San Diego Supercomputing Center (SDSC) through allocation PHY220161 from the Advanced Cyberinfrastructure Coordination Ecosystem: Services \& Support (ACCESS) program. We acknowledge support from the NSF grant DMR 1709781 and support from the Fisher General Endowment and SET grants from the Jess and Mildred Fisher College of Science and Mathematics at Towson University.

\end{acknowledgement}

\begin{suppinfo}

The supplementary information includes a histogram of the number of atoms in magnetic monolayer entries in C2DB, the generated monolayers, their magnetic moments, energies above the convex hull, and their prescence in OQMD, MP, C2DB, or JARVIS.

\end{suppinfo}

\bibliography{achemso-demo}
\includepdf[pages=-]{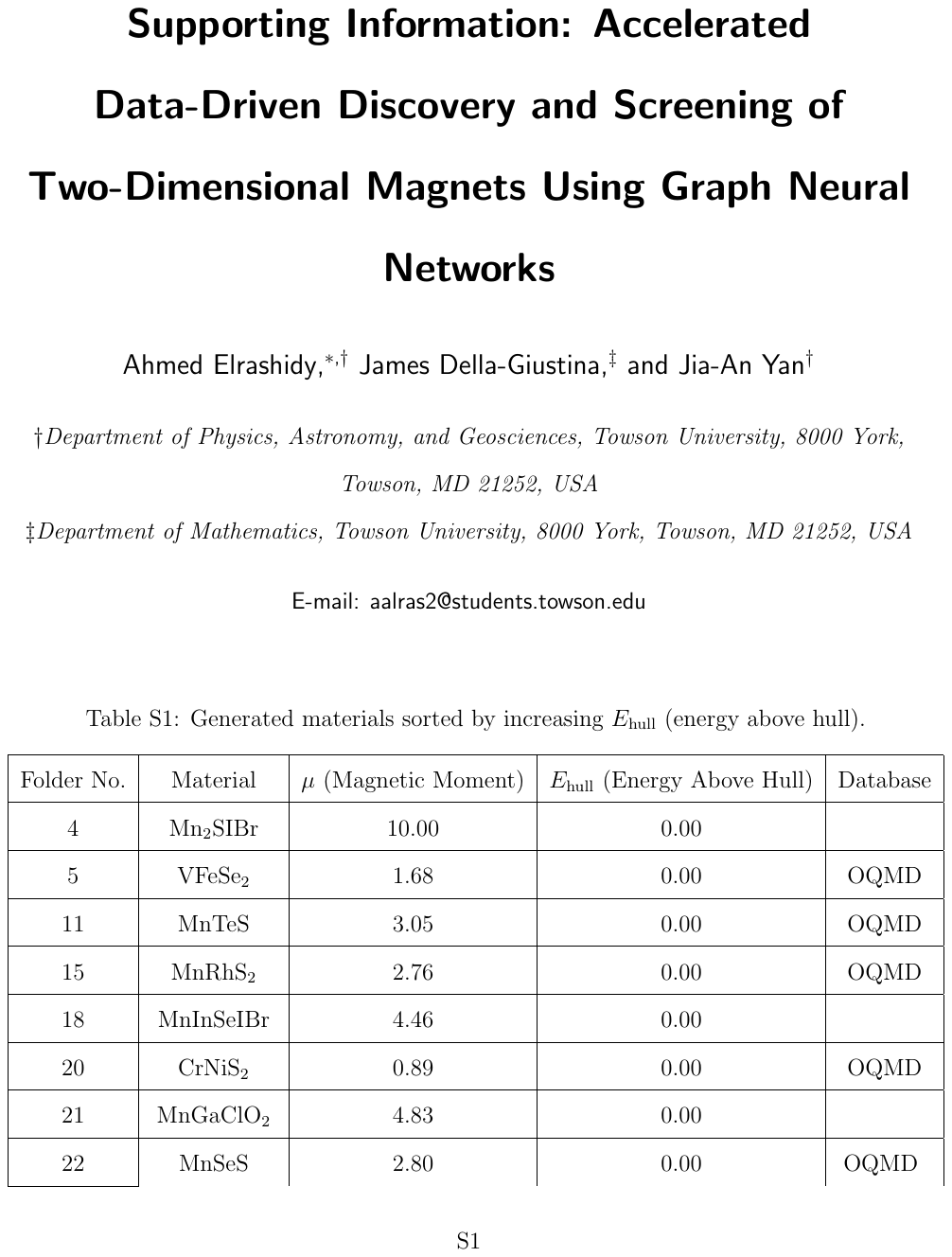}

\end{document}


\begin{longtable}{|c|c|c|c|c|}
\caption{Generated materials sorted by increasing $E_{\text{hull}}$ (energy above hull).}
\hline
Folder No. & Material & $\mu$ (Magnetic Moment) & $E_{\text{hull}}$ (Energy Above Hull) & Database \\ 
\hline
\endfirsthead
\hline
Folder No. & Material & $\mu$ (Magnetic Moment) & $E_{\text{hull}}$ (Energy above hull) & Database \\ 
\hline
\endhead
4 & Mn${_2}$SIBr & 10.00 & 0.00 & \\ \hline
5 & VFeSe${_2}$ & 1.68 & 0.00 & OQMD\\ \hline
11 & MnTeS & 3.05 & 0.00 & OQMD\\ \hline
15 & MnRhS${_2}$ & 2.76 & 0.00 & OQMD\\ \hline
18 & MnInSeIBr & 4.46 & 0.00 & \\ \hline
20 & CrNiS${_2}$ & 0.89 & 0.00 & OQMD\\ \hline
21 & MnGaClO${_2}$ & 4.83 & 0.00 & \\ \hline
22 & MnSeS & 2.80 & 0.00 & OQMD \hline
24 & MnInSe${_2}$I & 4.99 & 0.00 & \\ \hline
25 & MnBrClO & 3.00 & 0.00 & OQMD\\ \hline
26 & MnCl${_2}$O & 2.70 & 0.00 & OQMD \hline
31 & MnBrCl & 5.00 & 0.00 & OQMD\\ \hline
32 & MnI${_2}$O & 3.20 & 0.00 & OQMD\\ \hline
33 & CrIClO & 2.00 & 0.00 & \\ \hline
36 & MnClOF & 3.00 & 0.00 & \\ \hline
39 & MnGaS${_2}$Br & 5.00 & 0.00 & \\ \hline
41 & MnICl & 5.00 & 0.00 & OQMD\\ \hline
44 & VNCl${_2}$ & 0.03 & 0.00 & OQMD\\ \hline
45 & FeSeS & 1.66 & 0.00 & OQMD\\ \hline
49 & MnGaSeSI & 4.92 & 0.00 & \\ \hline
51 & NiHSO & 0.95 & 0.00 & OQMD\\ \hline
52 & GaNiO${_2}$F & 2.00 & 0.00 & \\ \hline
53 & NiHOF & 2.00 & 0.00 & OQMD\\ \hline
54 & MnCrSCl${_2}$ & 8.37 & 0.00 & \\ \hline
55 & MnTeSe & 2.81 & 0.00 & OQMD\\ \hline
58 & MnNiS${_2}$ & 2.97 & 0.00 & OQMD\\ \hline
59 & MnInS${_2}$Br & 5.00 & 0.00 & \\ \hline
60 & MnVTeSe & 1.32 & 0.00 & \\ \hline
61 & Fe${_2}$SeS & -0.00 & 0.00 & OQMD\\ \hline
62 & V${_2}$SBr${_2}$ & 6.00 & 0.00 & \\ \hline
63 & Mn${_2}$SBrCl & 10.00 & 0.00 & \\ \hline
64 & Mn${_2}$SICl & 10.00 & 0.00 & \\ \hline
68 & MnInSBr & 5.00 & 0.00 & \\ \hline
69 & MnSeSCl${_2}$ & 3.00 & 0.00 & \\ \hline
71 & CoHOF & 3.00 & 0.00 & OQMD\\ \hline
75 & MnNiSCl${_2}$ & 6.67 & 0.00 & \\ \hline
76 & FeBr${_2}$O & 4.12 & 0.00 & OQMD\\ \hline
77 & GaFeClO${_2}$ & 4.00 & 0.00 & \\ \hline
80 & MnCoS${_2}$ & 2.27 & 0.00 & OQMD\\ \hline
82 & MnVSeS & 1.74 & 0.00 & \\ \hline
85 & CrFeIClO & 6.93 & 0.00 & \\ \hline
87 & AlCrS${_2}$I & 3.98 & 0.00 & \\ \hline
88 & MnInSIBr & 4.48 & 0.00 & \\ \hline
89 & FeNiBr${_2}$O & 5.81 & 0.00 & \\ \hline
91 & VSIBr & 1.00 & 0.00 & \\ \hline
92 & MnInS${_2}$I & 4.98 & 0.00 & \\ \hline
93 & MnInSeSI & 5.00 & 0.00 & \\ \hline
94 & MnIClO & 3.53 & 0.00 & \\ \hline
95 & CrFeSICl & 6.97 & 0.00 & \\ \hline
99 & Mn${_2}$IBrO & 9.79 & 0.00 & \\ \hline
100 & MnVTeS & 3.19 & 0.00 & \\ \hline
102 & CrBrClO & 2.00 & 0.00 & \\ \hline
103 & FeHSO & 1.00 & 0.00 & OQMD\\ \hline
104 & MnCoSeS & 2.30 & 0.00 & \\ \hline
111 & VFeS${_2}$ & 0.99 & 0.00 & OQMD\\ \hline
114 & MnSbO${_3}$ & 4.00 & 0.00 & OQMD\\ \hline
116 & MnCrS${_2}$I & 8.00 & 0.00 & \\ \hline
118 & MnBr${_2}$O & 3.00 & 0.00 & OQMD\\ \hline
120 & CrNiSBrCl & 4.21 & 0.00 & \\ \hline
122 & MnAlS${_2}$Cl & 5.00 & 0.00 & \\ \hline
124 & CoBrCl & 3.00 & 0.00 & \\ \hline
125 & MnVTeCl${_2}$ & 7.88 & 0.00 & \\ \hline
127 & MnGaSeICl & 4.47 & 0.00 & \\ \hline
128 & Ti${_2}$SICl & 0.00 & 0.00 & \\ \hline
129 & MnGaBrO${_2}$ & 3.53 & 0.00 & \\ \hline
131 & VHClO & 3.00 & 0.00 & OQMD\\ \hline
132 & MnClF & 5.00 & 0.00 & OQMD\\ \hline
133 & MnAlS${_2}$ & 3.83 & 0.00 & OQMD\\ \hline
134 & Mn${_2}$TeSBr & 8.12 & 0.00 & \\ \hline
136 & MnPdSICl & 3.01 & 0.00 & \\ \hline
137 & MnVS${_2}$ & 0.97 & 0.00 & OQMD\\ \hline
138 & MnBiTeSeI & 5.00 & 0.00 & \\ \hline
140 & GaNiClO${_2}$ & 2.00 & 0.00 & \\ \hline
147 & Mn${_2}$SeSBr & 7.00 & 0.00 & \\ \hline
148 & VBrClO & 1.00 & 0.00 & OQMD\\ \hline
154 & VIBrO & 1.00 & 0.00 & \\ \hline
157 & MnOF${_2}$ & 2.78 & 0.00 & OQMD\\ \hline
158 & VHClO${_2}$ & 1.00 & 0.00 & \\ \hline
159 & ZrMnS${_2}$Br & 4.00 & 0.00 & \\ \hline
160 & Ti${_2}$SCl${_2}$ & -0.00 & 0.00 & \\ \hline
162 & ScMnSBr${_2}$ & 4.00 & 0.00 & \\ \hline
163 & Mn${_2}$SeBrCl & 10.00 & 0.00 & \\ \hline
164 & NbFeSeS & 1.47 & 0.00 & \\ \hline
172 & MnGaTeS${_2}$ & 3.80 & 0.00 & \\ \hline
175 & MnGaSe${_2}$Br & 4.84 & 0.00 & \\ \hline
177 & MnVSBrCl & 7.72 & 0.00 & \\ \hline
178 & MnVSBrCl & 7.72 & 0.00 & \\ \hline
179 & MnInSe${_2}$Cl & 5.00 & 0.00 & \\ \hline
180 & CrSeBr${_2}$ & 2.07 & 0.00 & \\ \hline
181 & NiF${_4}$ & 1.11 & 0.00 & \\ \hline
182 & MnInSeBr${_2}$ & 4.31 & 0.00 & \\ \hline
168 & CrSbSe${_2}$Br & 3.62 & 0.00 & \\ \hline
149 & CrFeS${_2}$ & 4.47 & 0.01 & OQMD\\ \hline
47 & Cr${_2}$SBr${_2}$ & 8.00 & 0.01 & \\ \hline
113 & MgMnSe${_2}$I & 4.00 & 0.01 & \\ \hline
35 & Ni${_2}$Br${_2}$O & 2.00 & 0.02 & OQMD\\ \hline
65 & CrGaTe${_3}$ & 3.01 & 0.02 & OQMD\\ \hline
115 & MnSeI & 3.21 & 0.02 & \\ \hline
155 & MnSBrCl & 3.02 & 0.02 & \\ \hline
142 & GaNiSBr${_2}$ & 1.07 & 0.03 & \\ \hline
108 & MnGeSeBr${_2}$ & 5.00 & 0.03 & \\ \hline
109 & MnCrS${_2}$ & 3.42 & 0.03 & OQMD\\ \hline
84 & CrSIBr & 2.22 & 0.03 & \\ \hline
9 & MnZnSI${_2}$ & 5.00 & 0.03 & \\ \hline
43 & MnNiSe${_2}$ & 3.21 & 0.03 & OQMD\\ \hline
42 & Mn${_2}$SeBr${_2}$ & 10.00 & 0.04 & \\ \hline
12 & Mn${_2}$SBr${_2}$ & 10.00 & 0.04 & \\ \hline
141 & MnSeICl & 3.14 & 0.04 & \\ \hline
97 & CrNiSe${_2}$ & 2.17 & 0.04 & OQMD\\ \hline
151 & CrCoSe${_2}$ & 1.03 & 0.05 & OQMD\\ \hline
17 & FeTeCl & 3.38 & 0.05 & OQMD\\ \hline
46 & VNiTe${_2}$ & 1.13 & 0.05 & OQMD\\ \hline
183 & FeCoSe${_2}$ & 3.41 & 0.05 & \\ \hline
169 & MnNiSBr${_2}$ & 6.92 & 0.05 & \\ \hline
171 & Mn${_2}$TeSI & 8.44 & 0.06 & \\ \hline
83 & MnCrTeSI & 8.00 & 0.06 & \\ \hline
96 & FeTeSe & 1.55 & 0.06 & OQMD\\ \hline
29 & NiIBr & 2.00 & 0.06 & OQMD\\ \hline
176 & Mn${_2}$Te${_2}$Se & 7.76 & 0.06 & \\ \hline
38 & MnZnS${_2}$I & 4.07 & 0.07 & \\ \hline
34 & MnSCl${_2}$ & 3.02 & 0.07 & \\ \hline
185 & CrSeI${_2}$ & 2.00 & 0.07 & \\ \hline
152 & Mn${_2}$Te${_3}$ & 7.92 & 0.08 & OQMD\\ \hline
112 & MnPdBr${_2}$ & 5.00 & 0.08 & \\ \hline
67 & Fe${_2}$TeS & 3.48 & 0.08 & \\ \hline
188 & Mn${_2}$SeICl & 9.69 & 0.08 & \\ \hline
48 & MnTe${_2}$Rh & 2.80 & 0.08 & OQMD\\ \hline
98 & MnNiTeSe & 3.00 & 0.09 & \\ \hline
86 & MnFeSeS & 4.08 & 0.09 & \\ \hline
57 & Mn${_2}$Se${_3}$ & 7.99 & 0.10 & OQMD\\ \hline
189 & CrCuS${_2}$I & 2.89 & 0.10 & \\ \hline
144 & FeCoTeSe & 3.51 & 0.10 & \\ \hline
126 & Fe${_2}$SCl${_2}$ & 7.97 & 0.10 & \\ \hline
56 & GaFeTeSeBr & 3.67 & 0.10 & \\ \hline
16 & MnCoSe${_2}$ & 5.26 & 0.11 & OQMD\\ \hline
173 & MnTe${_3}$Rh & 4.20 & 0.11 & \\ \hline
117 & Fe${_2}$TeSe & 5.04 & 0.11 & OQMD\\ \hline
7 & MnFeS${_2}$ & 4.66 & 0.12 & OQMD\\ \hline
0 & MnNiTe${_2}$ & 3.17 & 0.12 & OQMD\\ \hline
40 & MnFeSe${_2}$ & 3.43 & 0.12 & OQMD\\ \hline
110 & MnCrTe${_3}$ & 7.05 & 0.13 & OQMD\\ \hline
10 & VFeTe${_2}$ & 0.82 & 0.13 & OQMD\\ \hline
27 & MnFeTe${_2}$ & 3.96 & 0.14 & OQMD\\ \hline
161 & MnCrSe${_2}$ & 3.36 & 0.15 & \\ \hline
81 & Mn${_2}$GeTe${_2}$ & 3.70 & 0.15 & OQMD\\ \hline
106 & Mn${_2}$S${_3}$ & 7.87 & 0.15 & OQMD\\ \hline
135 & MnFeTe${_2}$S & 6.75 & 0.15 & OQMD\\ \hline
139 & MnGeSeI & 4.00 & 0.16 & \\ \hline
79 & CrTe${_2}$Rh & 2.42 & 0.16 & OQMD\\ \hline
2 & FeNiTe${_2}$ & 1.71 & 0.17 & OQMD\\ \hline
123 & MnVSe${_2}$ & 4.57 & 0.18 & OQMD\\ \hline
13 & Mn${_2}$TeS & 5.68 & 0.18 & OQMD\\ \hline
167 & Fe${_2}$GeTe${_2}$ & 4.23 & 0.18 & \\ \hline
105 & MnInSI & 5.00 & 0.18 & \\ \hline
37 & MnGeTeI & 4.00 & 0.20 & \\ \hline
14 & Mn${_2}$SeS & 9.71 & 0.20 & OQMD\\ \hline
74 & Mn${_2}$TeSe & 8.96 & 0.20 & OQMD\\ \hline
150 & MnFeS & 7.14 & 0.20 & \\ \hline
119 & MnInTe${_2}$S & 4.03 & 0.20 & \\ \hline
23 & MnGeI${_2}$ & 3.00 & 0.21 & \\ \hline
166 & VGeTe${_2}$ & 3.00 & 0.21 & \\ \hline
187 & Mn${_2}$SBr & 8.79 & 0.21 & \\ \hline
165 & MnPS${_2}$ & 4.17 & 0.22 & \\ \hline
107 & MnGeTeS & 3.00 & 0.22 & \\ \hline
121 & MnGeI & 3.91 & 0.22 & \\ \hline
30 & MnGe${_2}$ICl & 3.88 & 0.22 & \\ \hline
70 & MnGeCl & 3.51 & 0.22 & \\ \hline
90 & MnGeBr & 3.82 & 0.23 & \\ \hline
66 & FeCoTe${_2}$ & 2.70 & 0.25 & OQMD\\ \hline
73 & MnCrTe${_2}$ & 8.40 & 0.26 & OQMD\\ \hline
146 & Mn${_2}$TeSe${_2}$ & 4.12 & 0.26 & OQMD\\ \hline
143 & MnSbTeI & 4.37 & 0.40 & \\ \hline
1 & MnNiSeI${_2}$ & 3.59 & 0.59 & \\ \hline
\end{longtable}

\begin{figure}[H]
\includegraphics[scale = 0.7]{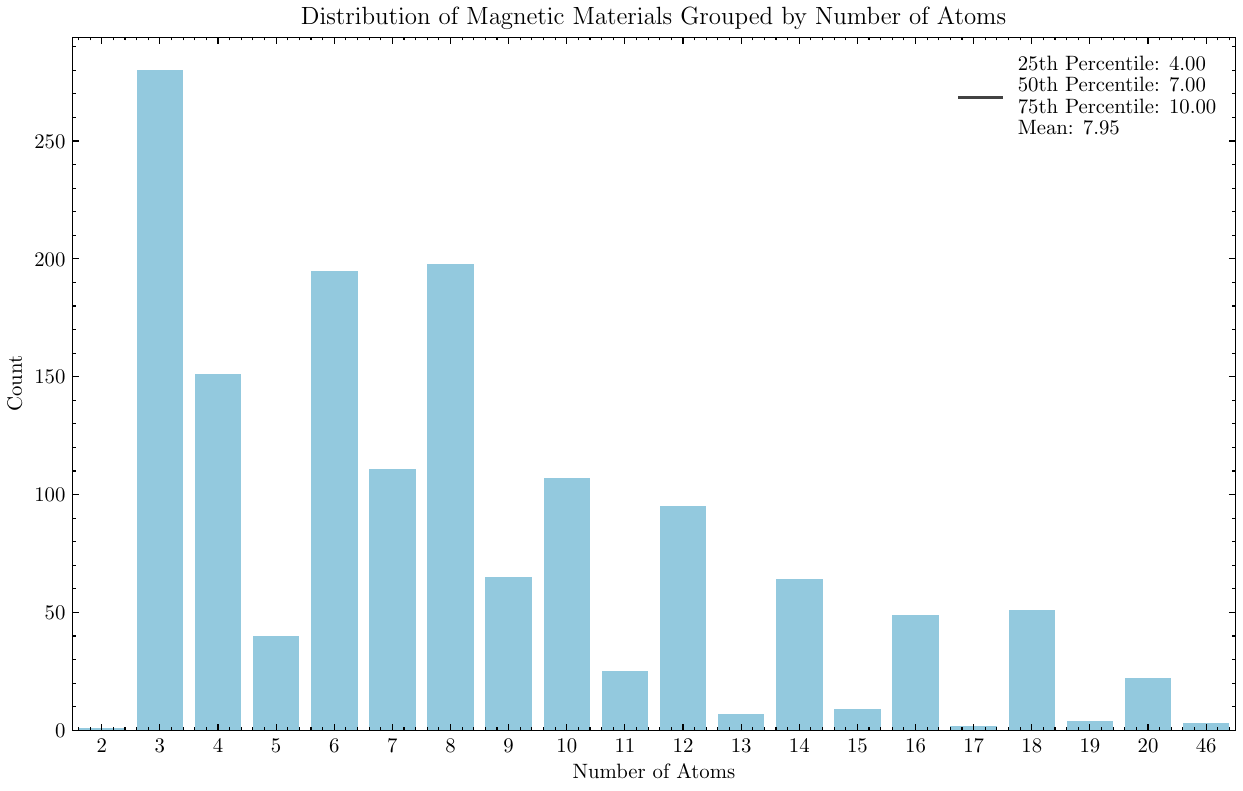}
\caption{A histogram of the number of atoms in magnetic monolayer entries in C2DB along with percentiles values.}
\label{num_atoms}
\end{figure}

\begin{figure}[H]
\includegraphics[scale = 0.7]{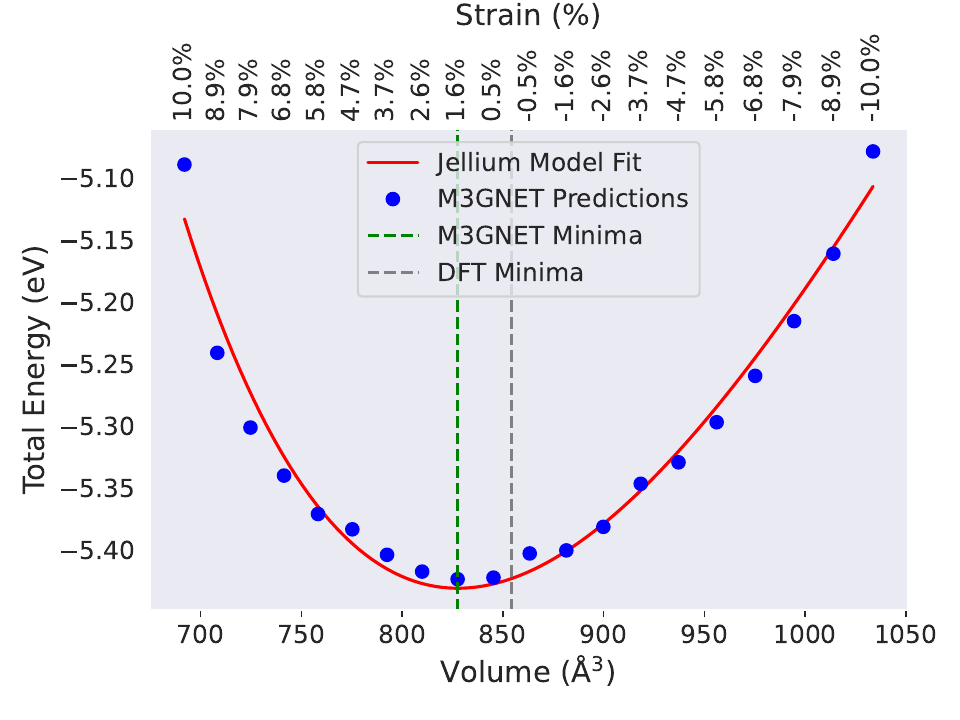}
\caption{Jellium model fit of the PES generated by ALIGNN-FF}
\label{align-ff}
\end{figure}